\documentclass[journal=aapmcd,manuscript=article]{achemso}
\usepackage{chemformula} % Formula subscripts using \ch{}
\usepackage[T1]{fontenc} % Use modern font encodings
\usepackage{braket}
\usepackage{hyperref}
\hypersetup{
    colorlinks=true,
    linkcolor=blue,
    urlcolor=blue,
    citecolor=blue,
}
\usepackage[capitalise]{cleveref}
\usepackage[euler]{textgreek}
\usepackage[version=4]{mhchem}

\author{Kin On Ho}
\altaffiliation{These authors contributed equally to this work.}
\author{Man Yin Leung}
\altaffiliation{These authors contributed equally to this work.}
\author{Yiu Yung Pang}
\altaffiliation{These authors contributed equally to this work.}
\author{King Cho Wong}
\author{Ping Him Ng}
\affiliation[The Chinese University of Hong Kong]
{Department of Physics, The Chinese University of Hong Kong,\\ Shatin, New Territories, Hong Kong, China}
\author{Sen Yang}
\email{syang@cuhk.edu.hk}
\affiliation[The Chinese University of Hong Kong]
{Department of Physics, The Chinese University of Hong Kong,\\ Shatin, New Territories, Hong Kong, China}
\alsoaffiliation[The Chinese University of Hong Kong, Shenzhen]
{Shenzhen Research Institute, The Chinese University of Hong Kong,\\ Shatin, New Territories, Hong Kong, China}

\title[ODMR x glue]
  {In-Situ Studies of Stress Environment in Amorphous Solids Using Negatively Charged Nitrogen Vacancy Centers in Nanodiamond}

\SectionNumbersOff

\begin{document}

%%%%%%%%%%%%%%%%%%%%%%%%%%%%%%%%%%%%%%%%%%%%%%%%%%%%%%%%%%%%%%%%%%%%%
%% The "tocentry" environment can be used to create an entry for the
%% graphical table of contents. It is given here as some journals
%% require that it is printed as part of the abstract page. It will
%% be automatically moved as appropriate.
%%%%%%%%%%%%%%%%%%%%%%%%%%%%%%%%%%%%%%%%%%%%%%%%%%%%%%%%%%%%%%%%%%%%%
%\begin{tocentry}

%\end{tocentry}

%%%%%%%%%%%%%%%%%%%%%%%%%%%%%%%%%%%%%%%%%%%%%%%%%%%%%%%%%%%%%%%%%%%%%
%% The abstract environment will automatically gobble the contents
%% if an abstract is not used by the target journal.
%%%%%%%%%%%%%%%%%%%%%%%%%%%%%%%%%%%%%%%%%%%%%%%%%%%%%%%%%%%%%%%%%%%%%
\begin{abstract}

Amorphous solids, which show characteristic differences from crystals, are common in daily usage. Glasses, gels, and polymers are familiar examples, and polymers are particularly important in terms of their role in construction and crafting. Previous studies have mainly focused on the bulk properties of polymeric products, and the local properties are less discussed. Here, we designed a distinctive protocol using the negatively charged nitrogen vacancy center in nanodiamond to study properties inside polymeric products in situ. Choosing the curing of poly dimethylsiloxane and the polymerization of cyanoacrylate as subjects of investigation, we measured the time dependence of local pressure and strain in the materials during the chemical processes. From the measurements, we were able to probe the local shear stress inside the two polymeric substances in situ. By regarding the surprisingly large shear stress as the internal tension, we attempted to provide a microscopic explanation for the ultimate tensile strength of a bulk solid. Our current methodology is applicable to any kind of transparent amorphous solids with the stress in the order of MPa and to the study of in situ properties in nanoscale. With better apparatus, we expect the limit can be pushed to sub-MPa scale.
\end{abstract}

%%%%%%%%%%%%%%%%%%%%%%%%%%%%%%%%%%%%%%%%%%%%%%%%%%%%%%%%%%%%%%%%%%%%%
%% Start the main part of the manuscript here.
%%%%%%%%%%%%%%%%%%%%%%%%%%%%%%%%%%%%%%%%%%%%%%%%%%%%%%%%%%%%%%%%%%%%%
\section{Introduction}

Amorphous solid, or non-crystalline solid, is a special class of materials that do not exhibit long-range order and do not possess a well-defined crystal structure due to the random nature of their solid formation. \cite{Stachurski2003, Shi2005, Stachurski2011} Glasses, gels, and polymers encountered on a daily basis are often amorphous. In particular, polymers make significant contributions to construction, manufacturing, and crafting industries. For instance, being optically clear and biocompatible, polydimethylsiloxane (PDMS) is widely used in various disciplines like lithography, medical devices, and contact lenses. Previous researches have revealed different bulk properties of PDMS. \cite{Khanafer2008, Saulnier2004, Johnston2014, Wang2016} Besides, adhesives (commonly known as glues) are also familiar examples of polymeric products, and they can bind individual materials together without deforming or damaging the adherends. Typically, adhesives are classified by adhesion mechanism first and then by reactivity. Among the reactive ones, Aron Alpha (AA) glue, with cyanoacrylate as the main component, is a one-part adhesive that provides long-lasting and strong stickiness for instant binding of materials. Many researches have investigated the bulk properties of adhesives in detail, for example, the mechanical \cite{Silva2006, Komurlu2016}, thermal, and magnetic properties \cite{Nakamura2018}.

One of the most frequently discussed properties of polymers is the internal stress development due to solidification or processing. This is a crucial topic for optimizing the manufacturing of polymeric products, such as preventing the formation of defects or cracks, and stabilizing the product shape or geometry. Macroscopic studies of the overall stress in polymer coatings and thin films can be readily found in the existing literature. Common measurement schemes include detecting the stress-induced curvature of the polymer-coated substrate, and analyzing the surface wrinkling of the polymer film when the underlying substrate is externally compressed. \cite{Francis2002, Chung2009} By invoking optical methods to observe the substrate curvature, the sensitivity and accuracy of measurement can be further improved. Undoubtedly, there are several mature and systematic methods to determine the internal stress as a bulk property of the whole polymeric material. On the other hand, probing the local stress environment is a relatively new research topic, and some local measurement schemes have been developed recently, such as fluorescent methods \cite{qiu2018, wang2014} and Raman spectroscopy. \cite{tang2013} Nonetheless, complicated procedures or sophisticated devices are sometimes needed to perform such measurements. To push forward this new type of research, we propose here an innovative protocol which uses the negatively charged nitrogen vacancy (NV$^{-}$) center in nanodiamond (ND) as a versatile sensor to study the local pressure and strain inside polymeric materials at a sub-micron scale. The distinctive ideas in our protocol include mapping the spatial heterogeneity in polymer with fine resolution, and performing \textit{in situ} measurements without the need of an elastic substrate to play a part in the stress analysis. In this work, PDMS and cyanoacrylate are chosen to be the sample polymeric materials.

Among the various point defects possibly found in diamonds, the NV center is a well-known one that offers a promising platform for realizing quantum technologies. The NV center consists of a substitutional nitrogen atom and an adjacent carbon vacancy. The presence of an extra electron turns NV into NV$^{-}$, which is a spin-1 system. Using optics for high-fidelity spin-state initialization and readout, reliable quantum sensing can be performed with the NV$^{-}$ center. In addition, modern technology enables the fabrication of ND with controllable size (possibly down to 10 nm). These NDs, each having a number of NV$^{-}$ centers inside, can serve as spatially resolved sensors with fine and adjustable resolution. In recent years, researches have revealed the exciting performance of the NV$^{-}$ center under stress environments \cite{Ho2020, Doherty2014P, Yip2019, Steele2017, Ivady2014, Broadway2019, Lesik2019, Hsieh2019}. Magnetic field sensing has also been successfully demonstrated under high pressure \cite{Yip2019, Lesik2019, Hsieh2019}. Furthermore, the sensing performance has been able to retain robustness over a wide temperature range \cite{Chen2011, Acosta2010, Doherty2014T, Ivady2014}. As a result, quantum sensing using the NV$^{-}$ center is an advantageous and effective approach to research topics that are, otherwise, problematic to deal with using classical sensors. Attempting to give new insights into the polymer research via quantum sensing, we applied our distinctive protocol to measure the time dependence and spatial distribution of local pressure and strain during the curing of PDMS and the polymerization of cyanoacrylate. Our microscopic measurement method can shed light on the local properties inside polymeric materials, as well as the dynamics of the curing and polymerization processes.

\begin{figure}
\includegraphics[width=8.5cm]{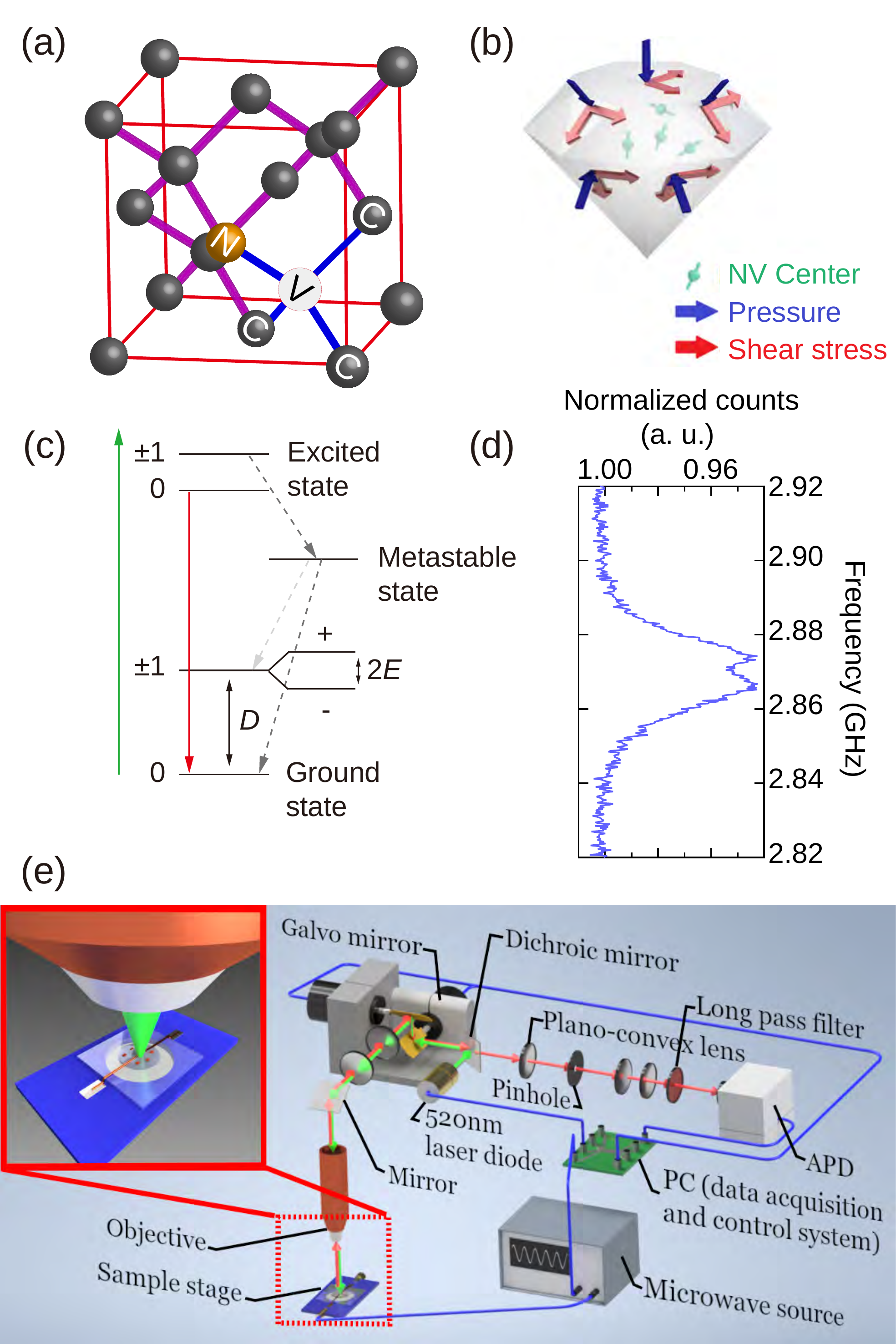}
\caption{(a) Structure of the NV$^{-}$ center. In a single crystal diamond, the nitrogen atom could substitute for any one of the four carbon atoms surrounding the vacancy, giving four possible NV$^{-}$ orientations relative to one of the four axes of the carbon bonding. (b) Schematic diagram about the possible distortions of a diamond crystal in a stress environment. Pressure is acted perpendicularly on the crystal surface, while shear stress is acted along the tangent planes of the crystal. Meanwhile, as indicated by the green-coloured symbols in the diagram, NV$^{-}$ centers in a diamond crystal have different orientations and spatial positions, leading to the inhomogeneous broadening of the ODMR peaks from a single ND. (c) Simplified energy structure of the NV$^{-}$ center. \textit{D} $\approx$ 2.87 GHz at ambient pressure and room temperature, and \textit{E} can vary from negligibly small to the order of MHz for different diamond samples. Besides, \textit{D} is related to pressure and temperature, while \textit{E} is related to strain and electric field. (d) An ODMR spectrum of a single ND is shown as an example. (e) Illustration of our confocal microscopy setup with a closeup of the sample stage. The spatial scanning is achieved by manoeuvring a galvo mirror. Using the confocal microscope, individual NDs can be probed.}
\label{fig1}
\end{figure}

Inside a single crystal diamond as shown in Fig.~\ref{fig1}(a), since the nitrogen atom could substitute for any one of the four carbon atoms surrounding the vacancy, the NV axis can be along any one of the four orientations of the carbon bonding. Besides, the strong crystal field around the NV$^{-}$ center is responsive to the deformation of the diamond lattice. From the schematic diagram in Fig.~\ref{fig1}(b), the diamond crystal can be distorted by pressure (normal forces) or shear stress (tangential forces). Depending on how the diamond is mechanically distorted, the energy levels of the NV$^{-}$ center inside would be altered differently. In Fig.~\ref{fig1}(c), the simplified energy structure of the NV$^{-}$ center is illustrated. By considering the first-order electron spin-spin interaction, the longitudinal zero-field splitting (ZFS) \textit{D} splits $\Ket{0}$ and $\Ket{\pm 1}$ states, while the transverse ZFS \textit{E} further splits the degenerate $\Ket{\pm 1}$ states into two superposition states of $\Ket{\pm 1}$ (eigenstates of $S_{x,y}$, labeled as $\Ket{+}$ and $\Ket{-}$) with an energy difference of 2\textit{E}. Here, the term \textit{D} is a function of pressure and temperature, whereas the term \textit{E} is a function of strain and electric field. A detailed discussion can be found in Ref. \cite{Barfuss2019, Mittiga2018}. Defining the NV axis as the z-axis, the corresponding ground-state Hamiltonian is given by
\begin{eqnarray}
H_{g} = D S_{z}^{2} + E \left( S_{x}^{2} - S_{y}^{2} \right) + \gamma \vec{B} \cdot \vec{S},
\label{eq:Hg}
\end{eqnarray}
where the longitudinal ZFS \textit{D} $\approx$ 2.87 GHz at ambient pressure and room temperature. With discrepancies between individual diamond samples, the transverse ZFS \textit{E} can vary from negligibly small to the order of MHz. The last term describes the Zeeman splitting under an applied magnetic field $\vec{B}$, where the gyromagnetic ratio $\gamma = 2.803$ MHz/G.

As shown in Fig.~\ref{fig1}(c), the spin-state dependent inter-system crossing of the NV$^{-}$ center enables the optical spin initialization and readout. With the spin-state dependent fluorescence rate, the electron spin resonance (ESR) can be measured by the optically detected magnetic resonance (ODMR) method \cite{Gruber1997}. An ODMR spectrum is shown in Fig.~\ref{fig1}(d) as an example. In the spectrum, the longitudinal ZFS \textit{D} and the traverse ZFS \textit{E} can be inferred from the center of resonances and the difference between resonances respectively. By measuring the changes in \textit{D}, \textit{E}, and the inhomogeneous broadening of the resonances, we can deduce the variations in the local stress environment. Moreover, the profile of magnetic field vectors can also be computed from the ODMR spectrum \cite{Yip2019}.

\section{Experimental details}
\subsection{Apparatus}

NV$^{-}$ center researches are usually accompanied by a confocal microscope with high resolution. An illustration of our home-built confocal microscope is shown in Fig.~\ref{fig1}(e). It has good fluorescence collection efficiency and spatial resolution, ensuring that ODMR measurements can be performed with high contrast. More detailed specifications of our setup are provided in the Supporting Information. On our sample stage, a large quantity of single-crystal NDs is spread throughout the surface of a glass slide. A 520 nm laser beam is focused onto the glass slide, across which a 50 \textmu m wire connecting to a microwave (MW) source is mounted. With the confocal microscope, the ODMR spectrum of NV$^{-}$ centers in an individual ND can be measured, where the green laser is used to initialize and readout the NV$^{-}$ centers' electron spin states, while the MW signal is responsible for manipulating the electron spin states. By means of ODMR spectroscopy, the variations in local pressure and strain at the positions of the single NDs can be probed.

One should be aware that temperature can produce a similar effect on the spectrum as pressure does. As temperature decreases, the value of \textit{D} increases. In fact, temperature and pressure effects can be treated separately. When pressure is kept constant, the temperature dependence of \textit{D} can be calibrated. The derivative $dD/dT$ takes the value of -74 kHz/K at room temperature and changes only slightly across a large temperature range \cite{Acosta2010, Chen2011, Doherty2014T}. To deal with the possible artifacts in the ODMR data brought by temperature fluctuations in the surroundings, we recorded the environment temperature continuously with a thermometer near the sample stage during all experiments. With reference to the first temperature datum, the environment temperature offset $\Delta T$ and hence the frequency offset $\Delta D = (dD/dT)\Delta T$ could be calculated for each ODMR measurement. By eliminating the change in \textit{D} due to the undesired temperature fluctuation in the environment, we isolated the spectral effects of pressure and local heating inside the sample. Besides, in our experiment with AA glue, two different ND sizes (1 \textmu m and 100 nm) were used to study the size effect introduced by the single NDs and demonstrate the adjustable spatial resolution of our sensing methodology. Refer to the Supporting Information for the characterization of our NDs. There is a variety of sensor sizes available in the market and one should chose according to the representative length scale in the physical system. Furthermore, power outputs of the MW source and the laser diode were fixed throughout each measurement. Nonetheless, the actual MW power and laser power received by the NDs might change with the ongoing chemical process in the sample during the experiment. These power effects could have altered the ODMR spectra of the NDs, but we checked that the possible variations in these powers would not be the primary factor leading to the detected changes in the measured quantities. See the Supporting Information for the full discussion of our data validity.

It has been reported that the surface termination of the NDs may influence the colloidal stability, distribution, and reactivity with the polymers \cite{Hajiali2017, Shakun2019, Zhao2019}. However, we designed our experimental protocol in a way that such undesired changes brought by the NDs to the system could be avoided. First, in each of our experiments, the density and total number of NDs were negligible as compared to the volume of the polymeric sample. This could be justified by the fact that our drop of pre-polymer was quite bulk and thick, while the NDs were hardly seen by naked eyes. A more rigorous evidence would be the sparse distribution of NDs observed in the confocal image as shown in Fig.~\ref{fig4}(e) and (f). Second, as stated on the website of Adámas Nanotechnologies (the supplier of our NDs), the surface chemistry of NDs mainly involves amphoteric surface functional groups (carboxylic acids, alcohols, etc.) with negative zeta potential. Since the same batch of ND is a type of non-reactive sensor for imaging a living system, together with the relatively low density of NDs, the curing process should not be significantly influenced by the sensors introduced to the system. Furthermore, as our NDs were on a glass slide rather than being mixed into the pre-polymer sample, we could reasonably consider that they had trivial effects, if any, on the bulk curing process. Nonetheless, one should keep in mind that it would be problematic if the pre-polymer was not bulk, or the density and total number of NDs were comparable to the volume of the pre-polymer.

\subsection{Sample preparation}

To prepare PDMS, we used a two-part liquid component kit called SYLGARD 184 silicone elastomer. The pre-polymer and curing agent were mixed in a ratio of 10:1 by stirring the two liquids thoroughly. To remove the bubbles formed, the mixture was allowed to degas in a vacuum chamber for an hour. After that, the PDMS mixture was ready to cure at room temperature.

For cyanoacrylate, the AA glue was used as the sample directly, without any special treatment. The polymerization of cyanoacrylate is triggered by the \ce{OH-} in ambient moisture, resulting in strong stickiness and hardness. This process is known to be an exothermic reaction. Nonetheless, the local heating from polymerization is assumed to be negligible here as the cyanoacrylates are mixed with solvents in AA glue. This assumption is validated by our data which show no sign of local temperature rise (see \textit{in situ} measurement for cyanoacrylate).

To implement the NDs as probes, we first got a bottle of NDs solution ready by adjusting the ND density and removing any sediments or clusters using ultrasound treatment. We then stuck a plasma-cleaned glass slide, on which the NDs solution was drop-casted, onto a specially designed printed circuit board (PCB). The electronic components for MW transmission were mounted at last. Before adding the sample onto the glass slide, all the measurement parameters were tested and confirmed, and a thermometer nearby was also turned on to start recording the environment temperature. During every measurement, the NDs were located at the interface between the glass slide and the sample to detect variations in the local stress environment of the polymeric substance. Unlike the elastic substrates in conventional stress studies, the glass slide in our sensing protocol was only to keep the sample \textit{in situ} and was not involved in any measurement procedures, allowing a direct microscopic analysis of the polymer \textit{in situ}. Refer to the Supporting Information for other remarks of the experiment.

We want to emphasis that in our current experimental design, the NDs are located at the interface between the glass slide and the amorphous solid. Due to the strong Van der Waals force, the NDs are unlikely to be taken up by the pre-polymer. This can also be understood by considering that no ND rotation is found in the magnetic field measurement as shown in the the Supporting Information. If the NDs were having significant movement inside the liquid of pre-polymer, it will be hard for our confocal microscope to track them. Meanwhile, mixing sensors into the sample is an interesting alternative approach by which we may extract more information from the measurements. Nonetheless, we first have to engineer the surface termination as well as develop a better tracking system, since the advantage of a confocal microscope is the high spatial resolution of the static objective, with the inability to track fast-moving objectives as a trade-off. Therefore, it would be more challenging in terms of sample preparation and optical detection if we were to mix the NDs into the pre-polymer.

\section{\textit{in situ} measurement for PDMS}

\begin{figure}
\includegraphics[width=8.5cm]{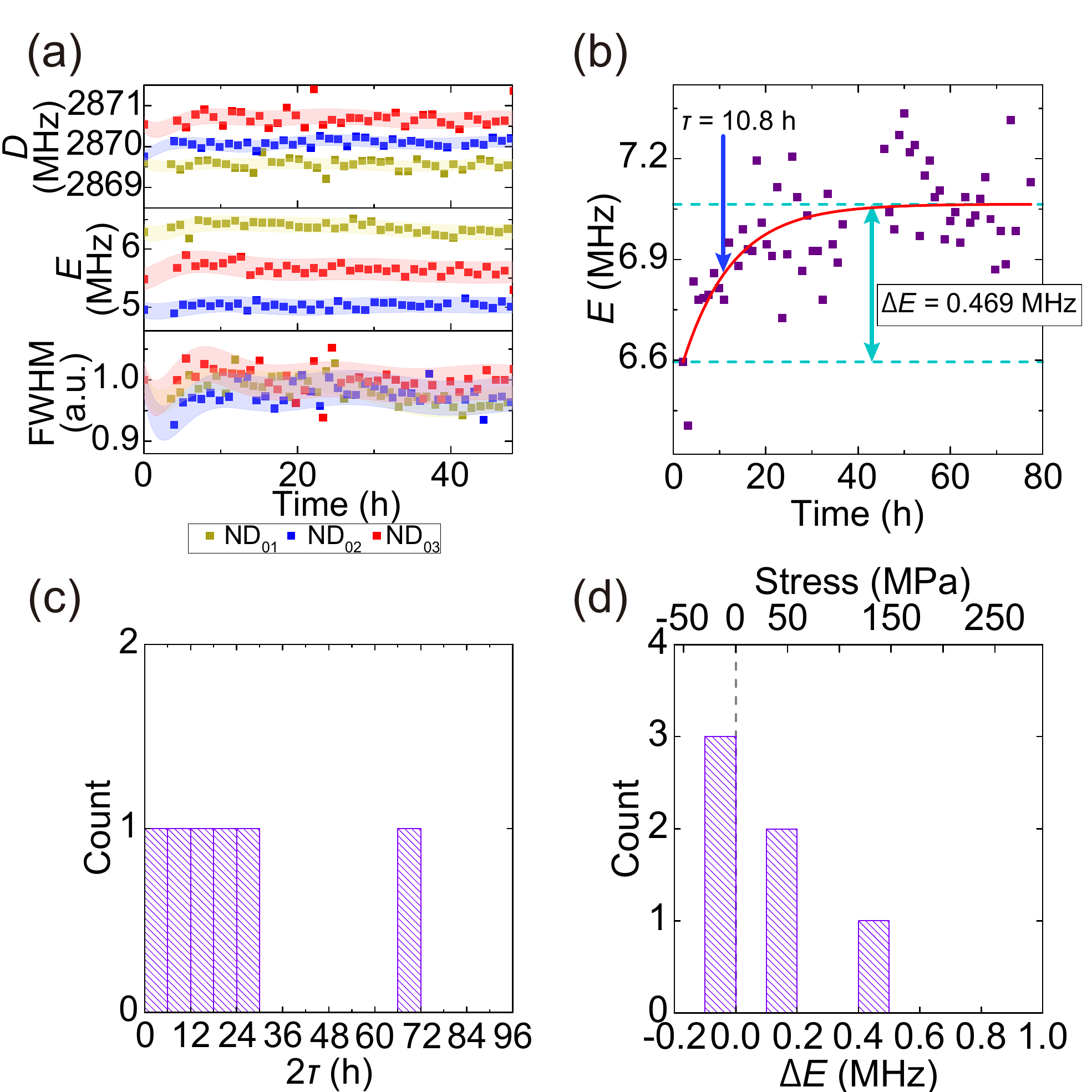}
\caption{(a) A measurement of the curing process of PDMS at room temperature using 1 \textmu m-NDs. The longitudinal ZFS \textit{D}, transverse ZFS \textit{E}, and average FWHM of the resonances are plotted against time. The first data point was always taken without PDMS. Besides, the FWHM data are presented as relative linewidths to the first data point in an attempt to eliminate the power broadening effect. In general, no observable change in \textit{D} and hence the pressure is present. However, a tiny increase in \textit{E} and FWHM is observed at the beginning, which may be due to the hardening of PDMS, while a tiny decrease in these two quantities is detected at a later time, which we suspect to be an outcome of the relaxation of strain in curing. (b) A sample fitting curve of the time dependence of \textit{E} using $y=y_{0}+a \cdot e^{-t/\tau}$, where this particular ND reveals that \textit{E} increases until saturation. From the fitting, $2\tau$ is defined as the measured curing time, while the last fitted value of \textit{E} minus the first measured value of \textit{E} after adding the PDMS is taken as the measured overall change in \textit{E} (\textDelta \textit{E}). (c, d) Two histograms respectively showing the statistical distributions of the curing times ($2\tau$) and the \textDelta \textit{E}s measured by the best 6 NDs in three separate measurements. A stress scale in MPa corresponding to the \textDelta \textit{E} scale in MHz is added to the top axis of (d), for relating the observable in NV physics to the actual magnitude of local shear stress in PDMS. See Discussion for calculation details.}
\label{fig2}
\end{figure}

1 \textmu m-NDs with nitrogen concentration of 3 ppm were used to measure the curing process of PDMS at room temperature, where each ND particle contains hundreds of thousands of NV$^{-}$ centers. A total of three separate measurements were performed. In each trial, from the confocal image of the PDMS sample (similar to that of a cyanoacrylate sample shown in Fig.~\ref{fig4}(e) and (f)), we selected a few NDs that were close to the MW transmission wire and showed high signal contrasts from the ODMR peaks. The targeted NDs were then used to track the variations of local properties inside the patch of PDMS. In one of these measurements, we chose 3 NDs to conduct ODMR spectroscopy repeatedly for more than 3 days, with no remarkable changes in the spectra after 48 hours. The measurement result up to 48 hours can be found in Fig.~\ref{fig2}(a), which concludes the temporal variations of the longitudinal ZFS \textit{D}, transverse ZFS \textit{E}, and average full width at half maximum (FWHM) of the two resonances. Here, the first data point was always taken before the addition of PDMS for a direct comparison of the quantities with and without the presence of PDMS. In particular, \textit{E} is likely to show an increase after adding the PDMS. From Fig.~\ref{fig2}(a), no observable change in \textit{D} and hence the pressure is detected, whereas a tiny increase in \textit{E} and FWHM is observed initially, which may be a consequence of the hardening of PDMS. We suspect that the curing process may occur on a random basis, leading to the development of strain inside PDMS. A tiny decrease in \textit{E} and FWHM is also generally measured at a later time. This detected strain relaxation may be explained by the viscoelasticity of PDMS. Since PDMS has a relatively low glass transition temperature ($T_g \approx$ -100 K) \cite{klonos2018}, our PDMS samples were always in the viscous state during the measurements in room conditions, and hence the PDMS chains were able to move around and rearrange themselves such that relaxation of the cross-linked polymer structure could happen. Meanwhile, oxidation of the viscoelastic PDMS samples under ambient conditions could lead to a hydrophobic recovery of PDMS. \cite{bodas2007} Our measured time dependence of strain is comparable to the result reported for another solidification mechanism of polymer. \cite{Francis2002} Nonetheless, the previously reported result was about the averaged stress in the entire polymeric material, whereas our result is showing microscopically the local strain inside PDMS. This is the main difference between our proposed protocol and traditional macroscopic methods for polymer research.

Although the same amorphous material and ND size were used in the three individual measurements, there are still subtle dissimilarities between their results. The first reason may be the difference in environment conditions where the experiments were conducted. The changes in ambient conditions such as humidity and temperature can affect the curing process. The second reason may be the difference in the patches of PDMS added onto the glass slide. Since we did not have precise control over the size, shape, and thickness of the PDMS sample, we could not exactly duplicate the patch of PDMS when we repeated the experiment. The third reason may be the difference in positions of the chosen NDs in the measurements. As the curing process can vary at different places in the PDMS, the data obtained may be position-dependent.

Besides, the ODMR data of the best 6 NDs from all three measurements were processed by fitting the time dependence of \textit{E} (excluding the data point before the addition of PDMS) with an exponential function $y=y_{0}+a \cdot e^{-t/\tau}$, where $y_{0}$, $a$, and $\tau$ are fitting parameters. Although the fitting example in Fig.~\ref{fig2}(b) shows a rising trend, \textit{E} may increase or decrease until reaching a stable value as measured by different NDs, and hence $a$ can be negative or positive. This possibly depends on whether or not that particular ND detects a significantly large local strain at the beginning of the curing process. With the fittings done, $2\tau$ is defined as the measured time taken for the curing process, while the last fitted value of \textit{E} minus the first measured value of \textit{E} after adding the PDMS is taken to be the measured overall change in \textit{E} (\textDelta \textit{E}) due to curing. To understand the statistical distributions of the curing times and the \textDelta \textit{E}s measured by the 6 chosen NDs, two histograms are plotted in Fig.~\ref{fig2}(c) and (d) respectively. The curing process generally lasts for less than a day at room temperature as revealed in Fig.~\ref{fig2}(c). On the other hand, by comparing the magnitudes of the positive and negative \textDelta \textit{E}s in Fig.~\ref{fig2}(d), the measured extent of the strain accumulation appears to be more prominent than that of the strain relaxation. To conclude, Fig.~\ref{fig2} successfully demonstrates the fineness in the sensing performance of NV$^{-}$ centers by showcasing tiny variations of the local properties in the PDMS. Besides, we attempted to compute the local shear stress inside the sample from changes in the ODMR spectrum. Refer to discussion for details.

\section{\textit{in situ} measurement for cyanoacrylate}
\begin{figure}
\includegraphics[width=8.5cm]{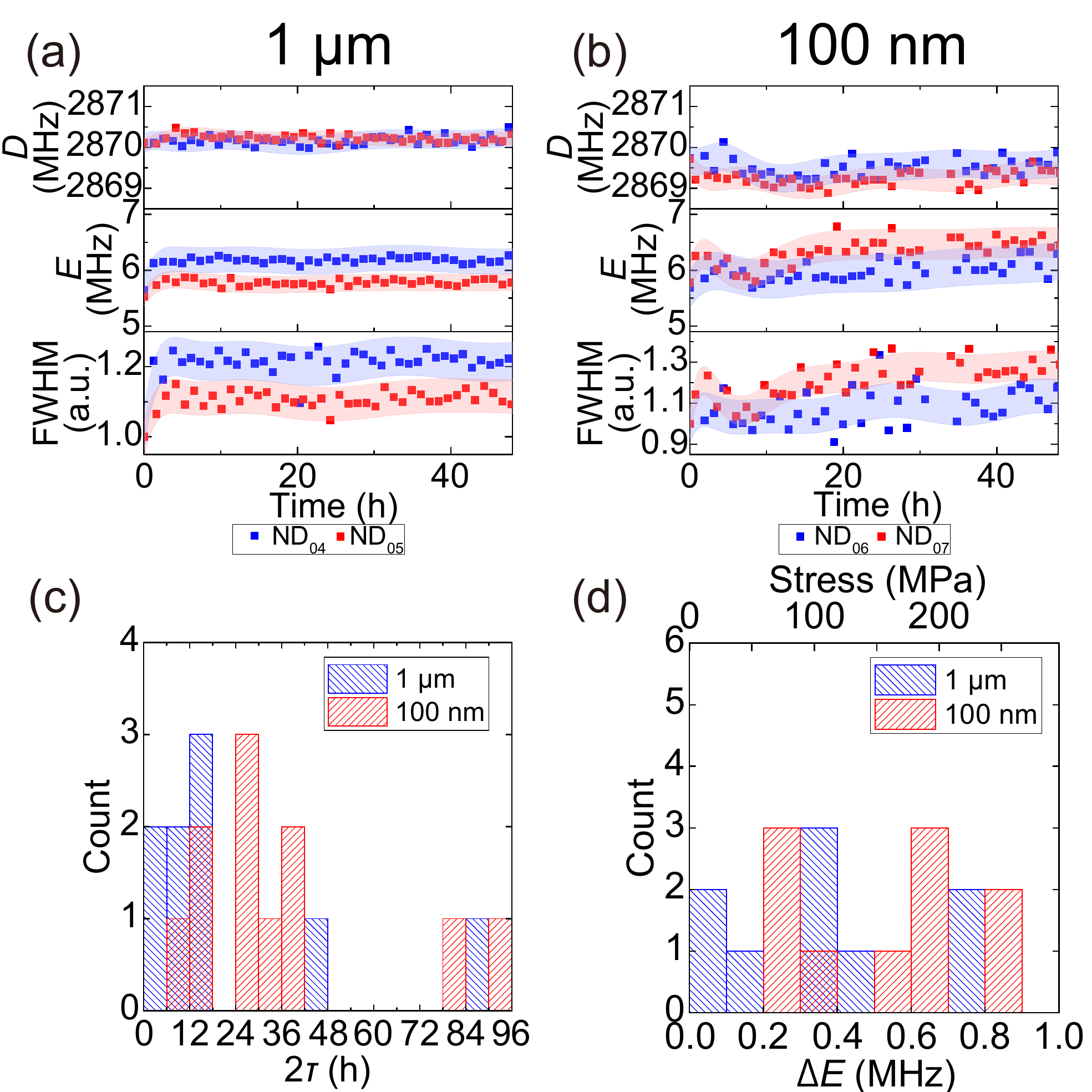}
\caption{(a, b) Two separate measurements of the polymerization process of cyanoacrylate at room temperature using 1 \textmu m-NDs and 100 nm-NDs respectively. Similar to PDMS, \textit{D}, \textit{E}, and average FWHM are plotted against time, with the first data point taken in the absence of AA glue and the FWHM data being relative values to the first datum. (a) There is no observable change in \textit{D} and hence the pressure. However, a slight increase in \textit{E} and FWHM is detected at the very beginning, and these two quantities remain stable until the end of measurement. (b) A spike in \textit{D}, \textit{E}, and FWHM is observed in the early stage of measurement, and \textit{D} remains stable afterwards, while \textit{E} and FWHM increase slowly in the remaining time, with a more noticeable increase shown in FWHM. (c, d) Two histograms respectively summarizing the measured polymerization times ($2\tau$) and the measured \textDelta \textit{E}s, each containing both the 1 \textmu m-ND and 100 nm-ND data sets. Following the same fitting procedures as in the last paragraph of the PDMS section, the polymerization time and \textDelta \textit{E} are defined in the same fashion here. Again, a stress scale in MPa is shown on the top axis of (d) to correlate the measured \textDelta \textit{E}s with the actual values of local shear stress inside AA glue. See Discussion for details. }
\label{fig3}
\end{figure}

Three separate 1 \textmu m-ND measurements and another four 100 nm-ND measurements were performed to study the polymerization of cyanoacrylate at room temperature, where both types of ND have the same nitrogen concentration of 3 ppm. The purpose of the experiments using 100 nm-NDs is to examine the size effect on the measured quantities. Again, in each of the seven trials, a few suitable NDs were used to monitor the variations of local properties inside the patch of AA glue. In both a 1 \textmu m-ND measurement and another 100 nm-ND measurement, we selected 2 NDs to conduct ODMR spectroscopy repetitively for more than 60 hours, with no significant changes in the spectra after 48 hours. The two individual measurement results up to 48 hours are displayed in Fig.~\ref{fig3}(a) and (b) respectively. Each of these figures summarizes the temporal variations of the longitudinal ZFS \textit{D}, transverse ZFS \textit{E}, and average FWHM of the two resonances. Similar to PDMS, the first data point was always taken without AA glue for a clear illustration of the glue's influence on the quantities, where \textit{E} generally increases compared to the data point before the addition of AA glue.

For the 1 \textmu m-ND measurement in Fig.~\ref{fig3}(a), the \textit{D} term has no observable change over time, implying that neither the pressure change nor the local heating is notable during the exothermic polymerization of cyanoacrylate (note that the possible artifacts from environment temperature fluctuations have been eliminated). On the other hand, the \textit{E} term and FWHM increase slightly in the first few hours and remain as a plateau afterwards. This suggests that local strain is accumulated in the early stage of measurement where the random polymerization readily occurs, and the local strain environment is stabilized at a later time when the polymerization is near completion. Moreover, the sensor position can affect the measured time taken for \textit{E} and FWHM to reach the plateau. Since the polymerization of cyanoacrylate is initiated by ambient moisture, the chemical process is expected to occur more rapidly on the edge and surface layer of the glue patch compared with the inner layers. Hence, being scattered in the glue sample, the NDs in the same experiment may disagree on when the local properties become stable. Only a microscopic sensing protocol like ours can detect such position dependence of the measured quantities and thus reflect the spatial heterogeneity of the solidification of AA glue.

For the 100 nm-ND measurement in Fig.~\ref{fig3}(b), information can be extracted more locally inside the AA glue with the smaller quantum sensors. In the figure, starting out with a spike in the first 10 to 20 hours, \textit{D} and hence the pressure show negligible changes afterwards, whereas \textit{E} and FWHM increase slightly until the end of the measurement, with a more pronounced increase in the FWHM data. The observation that \textit{E} and FWHM have an overall increasing trend indicates the presence and building up of strain in the AA glue during polymerization. We reckon that the discrepancy between the two measurements in Fig.~\ref{fig3}(a) and (b) can be explained by the effect of sensor size on the detection of local properties. Since diamond is the hardest material found in nature, it is expected to give tiny responses to small perturbations from a stress environment. Nonetheless, the sensitivity can be enhanced by reducing the size of the diamond, which means the smaller the diamond particle, the more sensitive it is. Thus, when subjected to weak external perturbations from the polymerization of cyanoacrylate, the 100 nm-ND would be a better sensor that reflects the local stress environment more accurately.

Besides, the ODMR data of the best 9 NDs in the three 1 \textmu m-ND measurements and the best 11 NDs in the four 100 nm-ND measurements were processed in the same manner as mentioned in the last paragraph of the PDMS section. With the same fitting function $y=y_{0}+a \cdot e^{-t/\tau}$, the measured time taken for polymerization and the measured overall change in \textit{E} (\textDelta \textit{E}) are defined likewise here. Nonetheless, different from the PDMS data, all the 20 chosen NDs in the AA glue experiments reveal similar increasing trends of \textit{E}, so the values of $a$ are always negative in the fittings while the values of \textDelta \textit{E} are all positive. To illustrate more clearly that the smaller sensors can have a higher sensitivity in measuring the polymerization time and \textDelta \textit{E}, a histogram superimposing the 1 \textmu m-ND and 100 nm-ND data sets is plotted for each of the quantities, as shown in Fig.~\ref{fig3}(c) and (d) respectively. By reducing the ND size, the peak in the statistical distribution of the measured polymerization times moves to the right, meaning that the smaller sensors generally detect a longer polymerization time than the larger ones. In fact, even if the glue, by direct observation, has already hardened and dried, residual reactions may persist on a small scale inside the glue patch. Thus, the smaller sensors can still capture the local variations of strain at a later stage of the experiment. In contrast, the larger sensors tend to reflect the overall chemical process occurring in the glue, as seen from the fact that most of the polymerization times measured by 1 \textmu m-NDs are within a day which is the usual time required for the hardening of AA glue. On the other hand, after switching to 100 nm-NDs, the statistical distribution of the measured \textDelta \textit{E}s shifts slightly to the right, indicating that a slightly larger increase in local strain owing to the random polymerization may be detected by the smaller sensors. This alteration in the statistical distribution may be attributed to the heterogeneity in the vitrified state of cyanoacrylate after polymerization. \cite{Stansbury2005} We speculate that the average spacing of the cross-linked polymers may be comparable to 100 nm. Hence, the 1 \textmu m-NDs, which are larger in size, may sense an averaged effect of the strain contributions from its surrounding polymer chains. This gives the possibility that, occasionally, the 100 nm-NDs would measure a slightly higher $\Delta E$ as compared to the 1 \textmu m ones. One may also wonder why both the 1 \textmu m-NDs and 100 nm-NDs show unremarkable changes in \textit{D} and hence the local pressure. We suspect that with shear strain as the main cause of the stickiness in AA glue, the normal forces on the ND surfaces are much weaker than the tangential forces during the polymerization, so the local pressure change is negligible in the measurement with either size of NDs. Moreover, we attempted to deduce the local shear stress accumulated in the glue from variations in the ODMR spectrum due to polymerization. Refer to discussion for details.

\begin{figure}
\includegraphics[width=8.5cm]{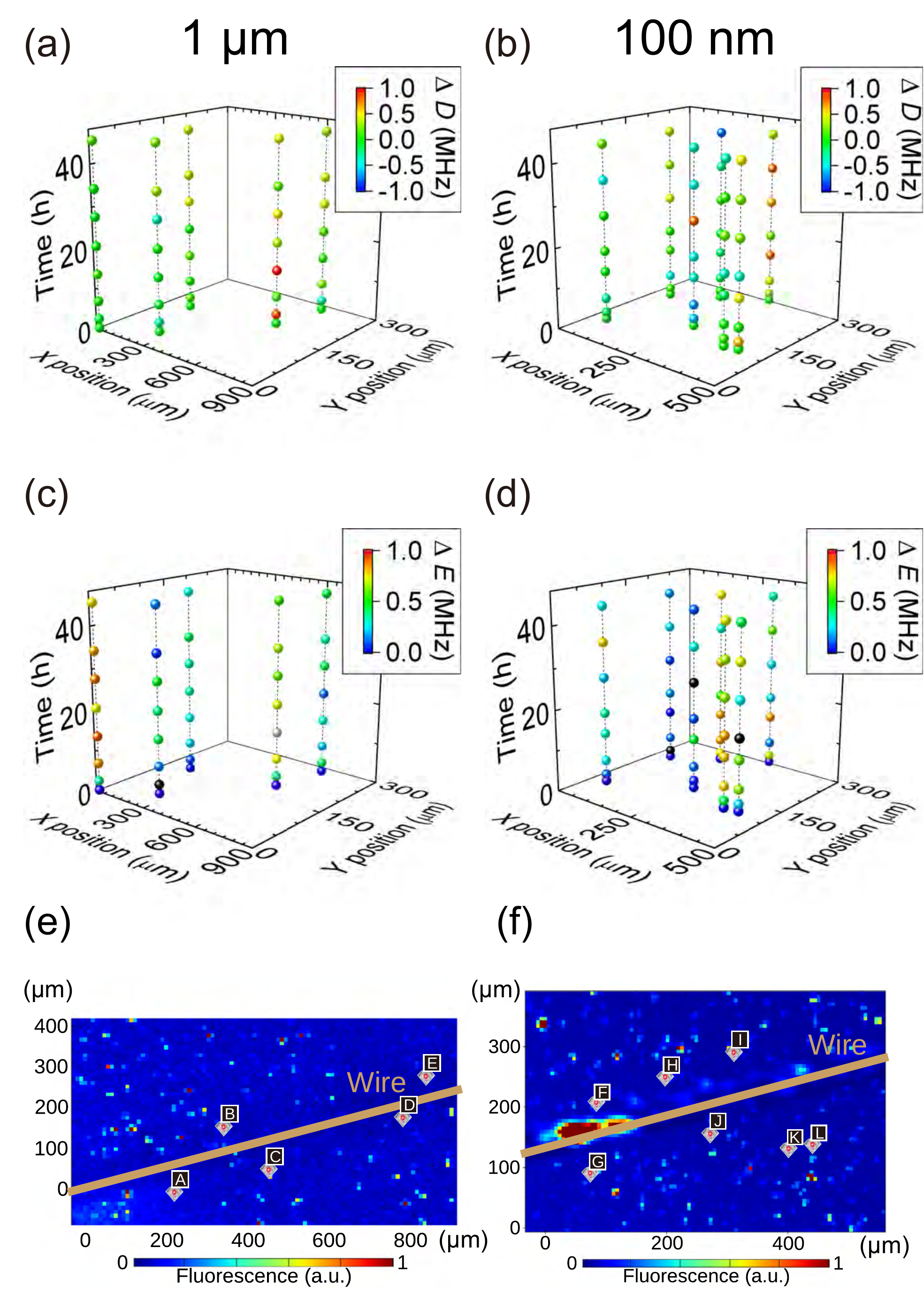}
\caption{(a, b) The spatial variations of $\Delta D$ over time in one 1 \textmu m-ND measurement and one 100 nm-ND measurement respectively. $\Delta D$ is defined as the change in \textit{D} from the first measured value after adding the glue. The $x$ and $y$ axes represent the $x$ and $y$ positions of a ND in the confocal image of the glue respectively, while the $z$ axis denotes the elapsed time in the experiment. Besides, the colour scale is for displaying the measured values of $\Delta D$, and a dashed line is used to connect the same ND positions at different times. During both experiments, tiny fluctuation in $\Delta D$ is observed at every ND position. (c, d) The spatial variations of $\Delta E$ over time in the two chosen measurements for (a) and (b) respectively. $\Delta E$ is defined as the change in \textit{E} compared to the first datum after adding the glue. The figures have the same format as those for $\Delta D$. In both measurements, the increase in \textit{E} varies at different ND positions. (e) The confocal image of the glue sample for the 1 \textmu m-ND measurement in (a) and (c). The fluorescence received by the photodetector is represented in terms of colour, with a reddish colour indicating a higher fluorescence level. The diamond symbols locate the chosen single-crystal NDs inside the glue, while the brownish yellow line shows the MW transmission wire on the glass slide. (f) The confocal image of the glue sample for the 100 nm-ND measurement in (b) and (d). The image format is the same as (e).}
\label{fig4}
\end{figure}

In addition to measuring the temporal variations of local properties, spatial mapping with fine resolution is also possible under our sensing scheme with NV$^{-}$ centers, which is a breakthrough for the polymer research community traditionally focusing on the overall stress in a bulk polymer. In Fig.~\ref{fig4}, $\Delta D$ and $\Delta E$ are respectively defined as the changes in \textit{D} and \textit{E} compared to their measured values just after adding the glue. For one 1 \textmu m-ND measurement and another 100 nm-ND measurement, the spatial variations of $\Delta D$ over time are illustrated in Fig.~\ref{fig4}(a) and (b) respectively. Likewise, Fig.~\ref{fig4}(c) and (d) are constructed with the $\Delta E$ data from the two chosen measurements respectively. In the figures, the ($x$,$y$) coordinates refer to the position of an ND in the confocal image of the glue sample (the confocal images for the 1 \textmu m-ND and 100 nm-ND measurements are displayed in Fig.~\ref{fig4}(e) and (f) respectively). The spatial distributions of $\Delta D$ and $\Delta E$ over time are consistent with some of our claims above. First, from Fig.~\ref{fig4}(a) and (b), the data points show a greenish colour in general, and the value of $\Delta D$ fluctuates slightly about zero at all the ND positions during the experiments. This supports our claim for the insignificance of the local pressure change during the polymerization of cyanoacrylate. Second, from Fig.~\ref{fig4}(c) and (d), although the NDs generally reveal an increasing trend of \textit{E}, the enhancement in \textit{E} varies at different ND positions. For example, in the 1 \textmu m-ND measurement, the ND with the smallest $x$ coordinate detects a relatively greater $\Delta E$ at the selected times, implying a stronger strain accumulation there. This agrees with our speculation that the rate and extent of polymerization are position-dependent in the glue sample, so the change in local strain induced by the reaction may differ from one position to another. Here, we have demonstrated the usage of NV$^{-}$ centers as spatially resolved sensors to map the spatial distribution of a physical quantity. In fact, we can even perform the ODMR measurement for an ND inside the glue under an external magnetic field, and we find that the targeted ND does not undergo rotation during the polymerization of cyanoacrylate. See the Supporting Information for details.

\section{Conclusions and Discussions}

In this work, we have demonstrated experimentally that the curing and polymerization processes can be studied using the NV$^{-}$ center in nanodiamond. Our measurement protocol reveals that the changes in pressure and shear stress can be measured \textit{in situ} with high resolution using 1 \textmu m-NDs and 100 nm-NDs. Our results provide hints regarding the local properties in the amorphous solids during the curing and polymerization processes. These experimental outcomes are beneficial to disciplines like soft condensed matter physics, biology, and industrial chemistry. Strictly speaking, our statistics are insufficient to make strong conclusions. In this paper, however, our main focus is to demonstrate the capability of using the NV centers in NDs as stress sensors. If considerably more measurements were made, we can quantitatively discuss the result. Nonetheless, in our current design of apparatus, it is time-consuming to do so since one measurement lasts for several days as mentioned. Therefore, we would like to focus on showing the proof-of-principle measurements.

Furthermore, using the average \textDelta \textit{E}s extracted from Fig.~\ref{fig2}(d) and Fig.~\ref{fig3}(d), we can approximately compute the local shear stresses induced by the curing of PDMS and the polymerization of cyanoacrylate respectively. In order to convert between the stress representations in NV physics and SI system, we referred to some previous theoretical calculations \cite{Barson2017, Broadway2019, Barfuss2019} and introduced the following additional assumptions. First, the axial (normal) forces on the surfaces of a ND are negligible, since our measurement results show insignificant changes in \textit{D} and hence the local pressure during the chemical processes. Second, the shear (tangential) forces on the surfaces of an ND are assumed, for simplicity, to have an equal probability of being in any particular direction, because we believe that the shear forces are random in direction inside our chosen materials. Indeed, it is possible to determine the components of the full stress tensor, but it requires a well-controlled source of external magnetic field. Third, we treat our system isotropically. Since we have no information on the angle between the NV axis and the stress direction, we cannot precisely determine the components of the shear stress. Nonetheless, the stress susceptibility parameters, $a_{1} = 4.86(2), a_{2} = -3.7(2), 2b = -2.3(3), \text{and } 2c = 3.5(3)$ in units of MHz/GPa \cite{Broadway2019}, do not differ by much in terms of magnitude. Therefore, for simplicity, we assume $a = b = c = 1.75$ MHz/GPa which means the ODMR response to the stress environment is isotropic. To precisely evaluate the local shear stress with respect to each axis, a well-controlled source of external magnetic field is, again, required. For PDMS, with the average \textDelta \textit{E} measured to be 0.1 MHz, the local shear stress induced by curing is calculated to be about 29.6 MPa. According to a bulk tensile test done by Johnston et al. \cite{Johnston2014}, the ultimate tensile strength (UTS) of PDMS is 5.13 $\pm$ 0.55 MPa at room temperature, thus the local shear stress obtained by our methodology is around 5.8 times larger. Likewise for AA glue, the average \textDelta \textit{E}s measured by 1 \textmu m- and 100 nm-NDs are about 0.35 and 0.5 MHz respectively, so the corresponding calculated values of local shear stress are about 104 and 148 MPa. As stated in the specification from the manufacturer, a bulk sample of an instant adhesive that is similar to our AA glue has a UTS of about 31.7 MPa, so the local shear stresses deduced from our 1 \textmu m-ND and 100 nm-ND data sets are about 3.3 and 4.7 times larger respectively. Here, both the PDMS and AA glue experiments reveal that the local accumulation of shear stress is quite strong. In fact, the local shear stress built up during the chemical processes can be considered as the internal tension in the amorphous solids. When experiencing a tensile load that exceeds the internal tension, the bulk solid starts fracturing. Thus, the local shear stress we obtained should be close to the UTS determined in bulk tensile tests. Nonetheless, our calculated local values are a few times greater than the bulk values. This difference may be attributed to the aforementioned assumptions, especially the approximation of an isotropic response from the ODMR spectrum to the stress environment. Meanwhile, we have observed a correlation between \textit{E} and FWHM in the ODMR spectrum, meaning that the change in \textit{E} reflects the variation in not only the local shear stress, but also the FWHM (decoherence time). Hence, interpreting the average \textDelta \textit{E} solely as the local shear stress accumulated would have been an over-estimation in our calculation. To conclude, despite being approximate values, we are able to microscopically probe the local shear stress inside PDMS and AA glue.

\section{Outlook}

The stress environment in amorphous solids is worthy of investigation due to its possible applications. As seen from the local shear stress detected during the curing of PDMS and the polymerization of cyanoacrylate, the solidification of amorphous substances may be easily used for applying shear stress to nanoparticles. Moreover, the high internal tension of adhesives may be beneficial for mechanical purposes. For example, adhesives are commonly used to attach a strain gauge to the testing material, so the strain gauge measurement can be affected by the characteristics of the adhesive, such as its thermal properties and the details of its solidifying reaction. \cite{Komurlu2016} Here, our methodology may provide more information on the microscopic behaviours of adhesives, enabling a more accurate usage of different adhesives as well as propelling the development of polymer science.

Besides, a better understanding of the curing process may help to refine the existing fabrication techniques of products, especially the products that will be in direct contact with the human body, such as contact lenses and medical equipments. On the other hand, polymerization is a crucial chemical reaction. In biology, RNA is one of the largest polymeric molecules in nature, and it is one of the most vital research topics related to the forms of life that has attracted not only the biologists, but also the physicists and chemists \cite{Spaeth2019, Schiessel2000, Higgs2016, Campbell2019, Costanzo2009, Rajamani2008, DaSilva2015, DeGuzman2014}. In industrial chemistry, dynamics of polymerization and various bulk properties of polymeric products are still active research areas \cite{Lin2019, Baldovin2019, Tomoshige2019, Nakamura2018, Sides2002, Silva2006, Komurlu2016, Grubbs2017}. Furthermore, beyond the field of amorphous solids and polymers, our methodology is also suitable for studying organic materials like perovskite solar cell. We believe that our proposed methodology would contribute to the material research field, facilitating the creation of various products that may improve the quality of life.

Apart from curing and polymerization, our suggested measurement protocol may be applied to investigate glass transition, which is driven by the change of thermodynamic variables such as temperature and pressure. Unlike polymerization where monomers join together to form polymers, glass transition describes the conversion from polymer to glass. Researches on this transition process are still growing and becoming increasingly important \cite{Gorham2019, Tomoshige2019, Zirdehi2019, Berthier2019, Luo2019, Drozd-Rzoska2007, Yuan2012, Fabbian1999}. In addition, we expect that, with careful design of instrument, it is possible to combine the optical setup with the standard tensile test setup. Therefore, the bulk and local tests can be performed simultaneously, giving comprehensive details on the materials' mechanical properties. \cite{Raja2016}
%%%%%%%%%%%%%%%%%%%%%%%%%%%%%%%%%%%%%%%%%%%%%%%%%%%%%%%%%%%%%%%%%%%%%
%% The "Acknowledgement" section can be given in all manuscript
%% classes.  This should be given within the "acknowledgement"
%% environment, which will make the correct section or running title.
%%%%%%%%%%%%%%%%%%%%%%%%%%%%%%%%%%%%%%%%%%%%%%%%%%%%%%%%%%%%%%%%%%%%%
\begin{acknowledgement}

The authors thank Chunshan Song for the fruitful discussion in chemistry. The authors thank King Yiu Yu for the fruitful discussions in the calculation. S.Y. acknowledges financial support from Hong Kong RGC (GRF/ECS/24304617, GRF/14304618 and GRF/14304419), CUHK start-up grant and the Direct Grants.

\end{acknowledgement}

%%%%%%%%%%%%%%%%%%%%%%%%%%%%%%%%%%%%%%%%%%%%%%%%%%%%%%%%%%%%%%%%%%%%%
%% The same is true for Supporting Information, which should use the
%% suppinfo environment.
%%%%%%%%%%%%%%%%%%%%%%%%%%%%%%%%%%%%%%%%%%%%%%%%%%%%%%%%%%%%%%%%%%%%%

\section{Associated content}

Details on our home-built confocal microscope, the characterization on the NDs, experimental and technical details of this work, including the data validity, and the measurement under magnetic field can be found.

%\begin{suppinfo}

%The following files are available free of charge.
%\begin{itemize}
%  \item Filename: brief description
%  \item Filename: brief description
%\end{itemize}

%\end{suppinfo}

%%%%%%%%%%%%%%%%%%%%%%%%%%%%%%%%%%%%%%%%%%%%%%%%%%%%%%%%%%%%%%%%%%%%%
%% The appropriate \bibliography command should be placed here.
%% Notice that the class file automatically sets \bibliographystyle
%% and also names the section correctly.
%%%%%%%%%%%%%%%%%%%%%%%%%%%%%%%%%%%%%%%%%%%%%%%%%%%%%%%%%%%%%%%%%%%%%
\bibliography{references}

\end{document}